\documentclass[aps,psfig,epsfig,preprint] {revtex4}
\usepackage{epsfig}
\usepackage{color}
\usepackage{bm}

\begin{document}
\draft
\title{
{\normalsize \hskip4.2in USTC-ICTS-06-06} \\{\bf  X(1576) as
Diquark-Antidiquark Bound State }}

\author{Gui-Jun Ding\footnote{e-mail address: dinggj@mail.ustc.edu.cn},
Mu-Lin Yan\footnote{e-mail address: mlyan@ustc.edu.cn}}

\affiliation{\centerline{Interdisciplinary Center for Theoretical
Study,} \centerline{University of Science and Technology of
China,Hefei, Anhui 230026, China} }

\begin{abstract}
We propose that the  broad $1^{--}$ resonance structure recently
discovered by BES in $J/\psi\rightarrow K^{+}K^{-}\pi^{0}$ is the
P-wave excitation of a diquark-antidiquark bound state. This
interpretation implies that there exists a negative parity, vector
nonet. A rough estimate of the mass spectrum of the nonet is
presented, and the prediction for the mass of $X(1576)$ is
consistent with the experimental data. The OZI allowed strong decays
are studied, it can decay into two pseudoscalars or one pseudoscalar
plus one vector meson. A crucial prediction is that $X(1576)$ should
dominantly decay into $K^{+}K^{-}$, $K_{L}K_{S}$, $\phi\pi^{0}$. The
observation of $I_3=1$ or $I_3=-1$ states which predominantly decays
into strange mesons could provide another important test to our
proposal. To search the charged $I_3=1$ isospin partner of
$X(1576)$, careful search in
$J/\psi\rightarrow K^{+}K_{L}\pi^{-}$, $J/\psi\rightarrow
K^{+}K_{S}\pi^{-}$ and $J/\psi\rightarrow \phi\pi^{+}\pi^{-}$ is
suggested.

\end{abstract}
\maketitle
\section{introduction}
A broad $1^{--}$ resonant structure X(1576) in $J/\psi\rightarrow
K^{+}K^{-}\pi^{0}$ has been reported by the BES collaboration
recently\cite{bes}. Its pole position is determined to be
$(1576^{+49+98}_{-55-91})$MeV-i$(409^{+11+32}_{-12-67})$MeV, and the
product branching ratio $Br(J/\psi\rightarrow
X(1576)\pi^{0})Br(X(1576)\rightarrow
K^{+}K^{-})=(8.5\pm0.6^{+2.7}_{-3.6})\times10^{-4}$. Therefore the
branching fraction of $J/\psi\rightarrow X(1576)\pi^{0}$ must be
much larger than $\mathcal{O}(10^{-4})$. Considering  the branching
ratio of the $J/\psi$ electromagnetic decay is usually of the order
$\mathcal{O}(10^{-4})$, so we determined that the decay
$J/\psi\rightarrow X(1576)\pi^{0}$ is mainly via the hadronic decay,
where both isospin and $G$-parity are conserved. Then X(1576) is of
even $G$-parity and its isospin $I=\mathbf{1}$, and the quantum
numbers of this structure are
$I^{G}(J^{PC})=1^{+}(1^{--})$\cite{gs,lipkin}. There is no obvious
standard $\rm{q}\bar{q}$ candidate for this state.

Since the decay products $K^{+}K^{-}$ contain a pair of strange
quark, it may contain a pair of hidden strange quark, and the
isospin triplet nature of this resonance requires that it at least
contain  additionally a pair of nonstrange quark, so it is
reasonable to expect that X(1576) is a diquark-antidiquark bound
state. The combined effects of the negative parity and the total
angular momentum $J=1$ require a unit of orbital angular momentum
excitation. Thus we are lead to the following assumption about the
structure of X(1576):
\begin{equation}
\label{1}X(1576)=\frac{1}{\sqrt{2}}(([ds][\bar{d}\bar{s}])_{\rm{P-wave}}-([su][\bar{s}\bar{u}])_{\rm{P-wave}})
\end{equation}

In the ref.\cite{maiani1} Maiani {\it et al.} pointed out that the
exotic states X(3872) and X(3940) can be well explained if they are
S-wave diquark-antidiquark bound state
$([cq][\bar{c}\bar{q}])_{\rm{S-wave}}$. Furthermore, they proposed
that the new state $Y(4260)$ may be the first orbital excitation of
a diquark-antidiquark bound state\cite{maiani2},
$Y(4260)=([cs][\bar{c}\bar{s}])_{\rm{P-wave}}$. If these are really
what happens in nature, it is reasonable to expect the P-wave
excitation of the four quark state
$([\rm{q}_1\rm{q}_2][\rm{\bar{q}}_3\rm{\bar{q}}_4])_{\rm{P-wave}}$($\rm{q}_i$
is light quark with i=1-4) should be seen experimentally, ${\it
i.e.,}$ the P-wave excitation of the nonet of light scalar
($J^{PC}=0^{++}$) mesons $\sigma_0(600), f_0(980), a(980),
\kappa(800)$. In our scheme, $X(1576)$ is exactly the P-wave
excitation of $a_0(980)$, and there exists analogously an nonet of
vector mesons with $J^{P}=1^{-}$. Henceforth, this nonet is denoted
by $X$. Since the width of $a_0$(980) is very large, the width of
X(1576) should also be large. Thus we qualitatively understood the
reason why the observed  width of X(1576) in the diquark-antiquark
picture is so huge. In this letter, we would like to give a rough
mass estimate of these states, and the prediction about the mass of
$X(1576)$ is consistent with its experimental value. The decay
properties of these states are discussed, which can decay into two
pseudoscalars or one pseudoscalar plus one vector meson, and some
distinctive predictions are given.

\section{mass spectrum of the vector nonet with $J^{PC}=1^{--}$}

\begin{figure}
\begin{center}
\includegraphics*[150pt,300pt][460pt,550pt]{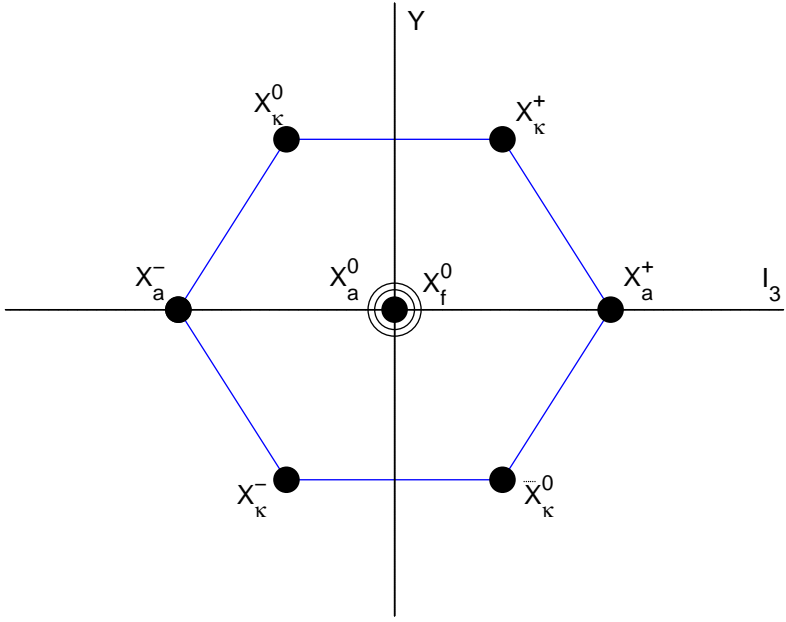}
\caption{The weight diagram of the nonet X}
\end{center}
\end{figure}

The weight diagram for the nonet is shown in fig.1, and we define
$[q_1q_2]\equiv\frac{1}{2}(q_1q_2-q_2q_1)$, then the composition of
the states of the nonet is as followings:
\begin{eqnarray}
\nonumber&&
X_a^+=([su][\bar{d}\bar{s}])_{\rm{P-wave}},~~~X_a^-=([ds][\bar{s}\bar{u}])_{\rm{P-wave}},~~~
X_a^0=\frac{1}{\sqrt{2}}(([ds][\bar{d}\bar{s}])_{\rm{P-wave}}-([su][\bar{s}\bar{u}])_{\rm{P-wave}}),\\
\nonumber&&
X_{\kappa}^{+}=([ud][\bar{d}\bar{s}])_{\rm{P-wave}},~~~X_{\kappa}^{-}=([ds][\bar{u}\bar{d}])_{\rm{P-wave}},~~~
X_{f}^{0}=\frac{1}{\sqrt{2}}(([ds][\bar{d}\bar{s}])_{\rm{P-wave}}+([su][\bar{s}\bar{u}])_{\rm{P-wave}}),\\
\label{2}&&
X_{\kappa}^{0}=([ud][\bar{s}\bar{u}])_{\rm{P-wave}},~~~\overline{X}_{\kappa}^{0}=([su][\bar{u}\bar{d}])_{\rm{P-wave}},~~~
X_{\sigma}^{0}=([ud][\bar{u}\bar{d}])_{\rm{P-wave}}
\end{eqnarray}
where for the two isosinglets, the states with definite strange
quark pair are introduced by assuming ideal mixing. The physical
states $X_{f}$ and $X_{\sigma}$ are mixing of $X^0_{f}$ and
$X^0_{\sigma}$ with mixing angle $\theta$,
\begin{equation}
\label{3}X_{f}=\cos\theta X^0_{f}+\sin\theta
X^0_{\sigma},~~X_{\sigma}=-\sin\theta X^0_{f}+\cos\theta
X^0_{\sigma}
\end{equation}

We will assume that the quarks prefer to form the "good" diquark
when possible. States dominated by that configuration should be
systematically lighter, more stable, and therefore more prominent
than the states formed from other types of diquarks. The residual
QCD interaction and the spin-orbit interaction will mix the $S=0$
"good" diquark with $S=1$ "bad" diquark( "good" and "bad" diquarks
in Jaffe's terminology\cite{jaffe1}), and a more sophisticated
treatment would have to consider these effects quantitatively.
However the effects only give a second order correction to the mass
and other properties, so we restrict to the "good" diquark in this
first analysis.

Most quark model treatments of multiquark spectroscopy use the
colormagnetic short range hyperfine interaction as the dominate
mechanism for possible binding\cite{colormag,lipkin2,sorba}. Here we
follow the same procedure, and the colormagnetic hyperfine
interaction is:
\begin{equation}
\label{4}H'=-\sum_{i>j}C_{ij}\vec{\lambda}_i\cdot\vec{\lambda}_i\;\vec{\sigma}_i\cdot\vec{\sigma}_j
\end{equation}
Here $\vec{\sigma}$ and $\vec{\lambda}$ are the Pauli and Gell-Mann
matrices, $i$ and $j$ run over the constituent quarks and
antiquarks. The coefficient $C_{ij}$ are dependent on the quark
masses and properties of the spatial wave functions of the quarks
and antiquarks in the system. In the SU(3) flavor symmetry limit,
$C_{ij}\equiv C$, and the standard treatment using the colorspin
$SU(6)_{cs}$ algebra gives the hyperfine energy
contribution\cite{jaffe2,hs2}:
\begin{equation}
\label{5} E'=\frac{C}{2}[D(tot)-2D(Q)-2D(\overline{Q})+16N]
\end{equation}
where $D=C_6-C_3-\frac{8}{3}S(S+1)$, and $D(tot)$, $D(Q)$,
$D(\overline{Q})$ denote the $D$ of the total system, the subsystem
of the quarks and the antiquarks respectively. $C_6$ and $C_3$ are
the quadratic Casimir operators of $SU(6)_{cs}$ and $SU(3)_c$
respectively, $S$ is the spin and $N$ is the total number of the
quarks and antiquarks. Rich phenomenology based on the colormagnetic
hyperfine interaction have been
developed\cite{sorba,hs1,hs2,jaffe2}, and a fit of charmed baryons
gives the consistent quark mass:
\begin{equation}
\label{6}m_{\rm{u}}\approx
m_{\rm{d}}\approx360\rm{MeV},~~~m_{\rm{s}}\approx540\rm{MeV},~~~m_{\rm{c}}\approx1710\rm{MeV}
\end{equation}
and the strength factors
\begin{equation}
\label{7}C_{\rm{qq}}=20\rm{MeV},~~~C_{qs}=12.5\rm{MeV},~~~C_{ss}=10\rm{MeV}
\end{equation}
Because the diquark  and antidiquark are in P-wave and are separated
by a distance larger than the range of the colormagnetic force, the
color hyperfine interaction operates only within the
diquark(antidiquark), but is not felt between the clusters.

There are three contributions to the mass of the states, {\it i.e.,}
the masses of the constituent quarks, the colormagnetic hyperfine
interaction energy, and the energy due to the P-wave excitation. We
estimate the contribution of the constituent quark mass from the
decay products($K^{+}K^{-}$), since their quark content is the same
as that of the the parent state($X(1576)$). Following the
Ref.\cite{maiani2}, the mass contribution due to the orbital angular
momentum can be estimated from the mass spectrum of the
$\rm{q\overline{q}}$ mesons with $L=0$ and $L=1$. The mass of the
$S=1$ states $K^{*}(892)$, $K_1(1270)$ and $K^{*}_2(1430)$ can be
described by the following equation:
\begin{equation}
\label{8}M(S,L,J)=K+2A\vec{S}\cdot\vec{L}+B\frac{L(L+1)}{2}
\end{equation}
where the second term is the spin-orbit interaction and the third
term is the mass contribution of the orbital angular momentum, Then
we find
\begin{equation}
\label{9}B=\frac{m_{K_1}+m_{K_2}-2m_{K^{*}}}{2}\approx458\rm{MeV}
\end{equation}
Then the mass contribution of the P-wave excitation for this set of
mesons equals $B$ which is approximately 458MeV. The mass of the
charm mesons $D^{*}(2007)^{0}$, $D_1(2420)^{0}$, $D^{*}_{2}(2460)^0$
can also be described by the formula Eq.(\ref{8}) with different
parameters $K, A, B$. In this case, the parameter $B$ is about 433
MeV, then the P-wave excitation energy for this set of charm mesons
approximately is 433 MeV. From above, we can see that the P-wave
excitation energy changes slowly with the meson mass variations
within the range about 1 $\rightarrow$ 2.5GeV. For simplicity, we
approximately take the P-wave excitation energy of the nonet $X$ to
be the average of the P-wave excitation energy of $K^{*}(892)$,
$K_1(1270)$, $K^{*}_2(1430)$ and that of $D^{*}(2007)^{0}$,
$D_1(2420)^{0}$, $D^{*}_{2}(2460)^0$, {\it i.e.,}
$E_{P}\approx\frac{458+433}{2}\approx445.5$MeV. By using  this
$E_P$, we determine the masses of $X$ as follows:
\begin{enumerate}

\item  The mass of $X_{a}(I=1)$ and $X^0_{f}$
\begin{equation}
\label{10}m_{X_{a}}=m_{X^0_{f}}=2(m_K+16C_{\rm{qs}})+E_P-8C_{\rm{qs}}-8C_{\rm{qs}}=2m_K+E_P+16C_{\rm{qs}}=1632.854\rm{MeV}
\end{equation}
This prediction is consistent with the experiment data, the pole
position of $X(1576)$ is:
$(1576^{+49+98}_{-55-91})$MeV-i$(409^{+11+32}_{-12-67})$MeV. Because
of the large decay width, it is very difficult to precisely
determine the mass of this resonance by experiments.

\item The mass of $X_{\kappa}$($X^{\pm}_{\kappa}$, $X^0_{\kappa}$ and
$\overline{X}^0_{\kappa}$)
\begin{equation}
\label{11}m_{X_{\kappa}}=m_{X_{a}}-8C_{\rm{qq}}+8C_{\rm{qs}}+m_{\rm{q}}-m_{\rm{s}}=m_{X_{a}}-240=1392.854\rm{MeV}
\end{equation}
If we use the experiment central value for the mass of $X_{a}$, the
peak mass of $X_{\kappa}$ is 1336 MeV.

\item The mass of $X^0_{\sigma}$
\begin{equation}
\label{12}m_{X^0_{\sigma}}=m_{X_{a}}-16C_{\rm{qq}}+16C_{\rm{qs}}+2m_{\rm{q}}-2m_{\rm{s}}=m_{X_{a}}-480=1152.354\rm{MeV}
\end{equation}
If the experimental value for the mass of $X_{a}$ is input, the peak
mass of $X^0_{\sigma}$ is 1096 MeV. The spectrum is similar to the
that of the light scalar nonet, which is inverted with respect to
the $\rm{q\bar{q}}$ nonet.
\end{enumerate}
\section{the decay of the vector nonet $X$}
The dominant decay mode of the four quark states is that they
dissociate into two colorless $\rm{q\bar{q}}$
mesons\cite{maiani2,jaffe2}, which means that a quark-antiquark pair
is switched between the diquark and antidiquark, then form a pair of
colorless $\rm{q\bar{q}}$ states. This mechanism has successfully
described the decay of the scalar nonet\cite{maiani3}, also has been
used to discuss the decay of other four quark states, and the
predictions for the decay width are close to the
experiment\cite{maiani1,maiani2,maiani3}. The nonet $X$ can decay
into two pseudoscalars or one pseudoscalar and one vector meson. In
the exact SU(3) flavor limit, the decay amplitude can be described
with a single parameter g, which describes the tunneling from the
bound diquark-antidiquark pair configuration to the meson-meson
pair. The parameters for the two pseudoscalars channel and one
pseudoscalar and one vector meson channel should be different, are
denoted as $g_1$ and $g_2$ respectively.
\subsection{$X\rightarrow$ pseudoscalar+ pseudoscalar }
We can describe the decay process by a single switch amplitude, {\it
e.g.,} the decay of $X^{+}_{a}$
\begin{equation}
\label{13}[su]_{\bar{3}_c}[\bar{d}\bar{s}]_{3_c}\rightarrow(s\bar{s})_{1_c}(u\bar{d})_{1_c}-(s\bar{d})_{1_c}(u\bar{s})_{1_c}
\end{equation}
where the subscripts indicate color configuration. Taking into
account the conservation of $C-$parity and $G-$parity, we can
further write the invariant three mesons effective coupling:
\begin{equation}
\label{14}ig_1X^{+\mu}_{a}[K^{-}\partial_{\mu}K^{0}-K^{0}\partial_{\mu}K^{-}]
\end{equation}
Here the coupling constant $g_1$ is dimensionless, and we will
introduce $\eta_{q}$ and $\eta_s$ in the following, which are
defined by
$\eta_q=\sqrt{\frac{2}{3}}\eta_1+\frac{1}{\sqrt{3}}\eta_8$,
$\eta_s=\frac{1}{\sqrt{3}}\eta_1-\sqrt{\frac{2}{3}}\eta_8$. The
physical states $\eta$, $\eta'$ are related to $\eta_8$ and $\eta_1$
via the usual mixing formula
$\eta_8=\eta\cos\theta_{p}-\eta'\sin\theta_{p}$,~$\eta_1=\eta\sin\theta_{p}+\eta'\cos\theta_{p}$
with the mixing angle $\theta_{p}=16.9^\circ\pm1.7^\circ$\cite{mix}.
From the effective lagrangian (\ref{14}), we find the decay
amplitude:
\begin{eqnarray}
\label{15}{\cal M}(X^{+}_{a}\rightarrow
K^{+}\overline{K}^{0})&=&g_1\;\varepsilon^{\mu}(X^+_a)[p_{\mu}(K^+)-p_{\mu}(\overline{K}^0)]
\end{eqnarray}
Here $\varepsilon^{\mu}(X^+_a)$ is the polarization vector of
$X^{+}_{a}$, $p_{\mu}(K^+)$ is the four momentum vector of $K^{+}$.
The decay of the other member of the nonet can be investigated in
the same way, and the effective lagrangian for the relevant decays
is as followings,
\begin{eqnarray}
\nonumber{\cal
L}_{eff}&=&ig_1\{X^{+\mu}_{a}[K^{-}\partial_{\mu}K^{0}-K^{0}\partial_{\mu}
K^{-}]+X^{-\mu}_{a}[-K^{+}\partial_{\mu}\overline{K}^{0}+\overline{K}^{0}\partial_{\mu}
K^{+}]\\
\nonumber&&+X^{+\mu}_{\kappa}[\overline{K}^{0}\partial_{\mu}\pi^{-}-\pi^{-}\partial_{\mu}\overline{K}^{0}+\frac{1}{\sqrt{2}}K^{-}\partial_{\mu}(-\pi^0+\eta_q)-\frac{1}{\sqrt{2}}(-\pi^0+\eta_q)\partial_{\mu}
K^{-}]\\
\nonumber&&+X^{-\mu}_{\kappa}[-K^{0}\partial_{\mu}\pi^{+}+\pi^{+}\partial_{\mu}
K^{0}-\frac{1}{\sqrt{2}}K^{+}\partial_{\mu}(-\pi^0+\eta_q)+\frac{1}{\sqrt{2}}(-\pi^0+\eta_q)\partial_{\mu}
K^{+}]\\
\nonumber&&+X^{0\mu}_{\kappa}[-K^{-}\partial_{\mu}\pi^{+}+\pi^{+}\partial_{\mu}K^{-}+\frac{1}{\sqrt{2}}(\pi^{0}+\eta_q)\partial_{\mu}\overline{K}^0-\frac{1}{\sqrt{2}}\overline{K}^0\partial_{\mu}(\pi^{0}+\eta_q)]\\
\nonumber&&+\overline{X}^{0\mu}_{\kappa}[K^{+}\partial_{\mu}\pi^{-}-\pi^{-}\partial_{\mu}K^{+}-\frac{1}{\sqrt{2}}(\pi^{0}+\eta_q)\partial_{\mu}K^0+\frac{1}{\sqrt{2}}K^0\partial_{\mu}(\pi^{0}+\eta_q)]\\
\nonumber&&+X^{0\mu}_{a}\frac{1}{\sqrt{2}}[K^{+}\partial_{\mu}K^{-}-K^{-}\partial_{\mu}K^{+}+K_{L}\partial_{\mu}K_{S}-K_{S}\partial_{\mu}K_{L}]\\
\label{16}&&+X^{0\mu}_{f}\frac{1}{\sqrt{2}}[-K^{+}\partial_{\mu}K^{-}+K^{-}\partial_{\mu}K^{+}+K_{L}\partial_{\mu}K_{S}-K_{S}\partial_{\mu}K_{L}]\}
\end{eqnarray}
From the above lagrangian, we can calculate the width of various
decay channels following standard procedure. The decay width is
expressed as
\begin{equation}
\label{17}\Gamma(X\rightarrow P_1+P_2)=\frac{g^2_1}{6\pi
}C_{X\rightarrow P_1P_2}
\frac{2}{(1-\beta)\sqrt{2\pi}\;\Gamma_X}\int^{m_X+\delta}_{m_X-\delta}dm\;\frac{|\vec{p}|^3}{m^2}\exp[-\frac{(m-m_X)^2}{2(\Gamma_X/2)^2}]
\end{equation}
where because of the large decay width, the mass distribution has
been considered by using a exponential function\cite{pdg}.
$|\vec{p}|$ is the decay momentum
$|\vec{p}|=\frac{\sqrt{(m^2-(m_1+m_2)^2)(m^2-(m_1-m_2)^2)}}{2m}$,
$\delta=1.64\frac{\Gamma_X}{2}$, $\beta=10\%$\cite{pdg}. $m_1$ and
$m_2$ are respectively the mass of two pseudoscalars $P_1$ and
$P_2$. $C_{X\rightarrow P_1P_2}$ is a numerical coefficient which
can be found from the effective lagrangian (\ref{16}), and
coefficient $C_{X\rightarrow P_1P_2}$ for various decays are listed
in Table I.
\begin{table}[hptb]
\begin{center}
\caption{the numerical coefficient entering Eq.(\ref{17}) for the
decay of the nonet X}
\begin{tabular}{|c|c|c|c|}\hline\hline
Decay  &   $C_{X\rightarrow P_1P_2}$  &  Decay  &  $C_{X\rightarrow
P_1P_2}$\\ \hline

$X^{+}_a\rightarrow K^{+}\overline{K}^{0}$ & 1
&$X^{+}_{\kappa}\rightarrow \pi^{+}K^{0}$&$1$
\\ \hline

$X^{+}_{\kappa}\rightarrow
K^{+}\pi^{0}$&$\frac{1}{2}$&$X^{+}_{\kappa}\rightarrow
K^{+}\eta$&$\frac{1}{2}(\sqrt{\frac{2}{3}}\sin\theta_{p}+\sqrt{\frac{1}{3}}\cos\theta_p)^2$\\
\hline

$X^{+}_{\kappa}\rightarrow
K^{+}\eta'$&$\frac{1}{2}(\sqrt{\frac{2}{3}}\cos\theta_{p}-\sqrt{\frac{1}{3}}\sin\theta_p)^2$&$X^{0}_{\kappa}\rightarrow\pi^{-}K^{+}$&1\\
\hline

$X^{0}_{\kappa}\rightarrow\pi^{0}K^{0}$&$\frac{1}{2}$&$X^{0}_{\kappa}\rightarrow
K^{0}\eta$&$\frac{1}{2}(\sqrt{\frac{2}{3}}\sin\theta_p+\sqrt{\frac{1}{3}}\cos\theta_p)^2$\\
\hline

$X^{0}_{\kappa}\rightarrow
K^{0}\eta'$&$\frac{1}{2}(\sqrt{\frac{2}{3}}\cos\theta_p-\sqrt{\frac{1}{3}}\sin\theta_p)^2$&$X^{0}_{a}\rightarrow
K^{+}K^{-}$&$\frac{1}{2}$\\ \hline

$X^{0}_{a}\rightarrow
K_{L}K_{S}$&$\frac{1}{2}$&$X^{0}_{f}\rightarrow K^{+}K^{-}$&$\frac{1}{2}$\\
\hline

$X^{0}_{f}\rightarrow
K_{L}K_{S}$&$\frac{1}{2}$& &\\
\hline\hline
\end{tabular}
\end{center}
\end{table}
In this table, we have not shown the decay channels which can be
obtained from the channels appearing in the table by making charge
conjugation, {\it e.g.,} for $X^{-}_a\rightarrow K^{-}K^{0}$, the
corresponding numerical coefficient is $\mathbf{1}$. From Table I,
we can see that the dominant decay modes of $X^{0}_a$($X(1576)$) are
$K^{+}K^{-}$ and $K_{L}K_{S}$, and
\begin{equation}
\label{18}\frac{\Gamma(X^{0}_a(X(1576))\rightarrow
K^{+}K^{-})}{\Gamma(X^{0}_a(X(1576))\rightarrow K_{L}K_{S})}\approx1
\end{equation}
However, $X^{0}_a$($X(1576)$) can not decay into $\pi^{+}\pi^{-}$,
\begin{equation}
\label{19}\Gamma(X^{0}_a(X(1576))\rightarrow \pi^{+}\pi^{-})\approx0
\end{equation}
Dominant $K^{+}K^{-}$ and $K_{L}K_{S}$ decays is a distinctive
signature of the validity of the present model. Some interesting
relations can be found, such as:
\begin{eqnarray}
\nonumber&&\Gamma(X^{+}_{a}\rightarrow
K^{+}K_{L})\approx\Gamma(X^{+}_{a}\rightarrow
K^{+}K_{S})\approx\Gamma(X^{0}_{a}\rightarrow K^{+}K^{-})\approx\Gamma(X^{0}_{a}\rightarrow K_{L}K_{S})\\
\nonumber&&\Gamma(X^{0}_{f}\rightarrow
K^{+}K^{-})\approx\Gamma(X^{0}_{f}\rightarrow
K_{L}K_{S})\approx\Gamma(X^{0}_{a}\rightarrow K^{+}K^{-})\\
\nonumber&&\Gamma(X^{+}_{\kappa}\rightarrow\pi^{+}K^{0})\approx2\Gamma(X^{+}_{\kappa}\rightarrow
K^{+}\pi^{0})\approx\Gamma(X^{0}_{\kappa}\rightarrow K^{+}\pi^{-})\approx2\Gamma(X^{0}_{\kappa}\rightarrow K^{0}\pi^{0})\\
\nonumber&&\Gamma(X^{+}_{\kappa}\rightarrow
K^{+}\eta)\approx\Gamma(X^{0}_{\kappa}\rightarrow
K^{0}\eta),~\Gamma(X^{+}_{\kappa}\rightarrow
K^{+}\eta')\approx\Gamma(X^{0}_{\kappa}\rightarrow K^{0}\eta')\\
\nonumber&&\tilde{\Gamma}(X^{+}_{\kappa}\rightarrow\pi^{+}K^{0})=\tilde{\Gamma}(X^{+}_{\kappa}\rightarrow
K^{+}\pi^{0})+\tilde{\Gamma}(X^{+}_{\kappa}\rightarrow
K^{+}\eta)+\tilde{\Gamma}(X^{+}_{\kappa}\rightarrow K^{+}\eta')\\
\label{20}&&\tilde{\Gamma}(X^{0}_{\kappa}\rightarrow\pi^{-}K^{+})=\tilde{\Gamma}(X^{0}_{\kappa}\rightarrow\pi^{0}K^{0})+\tilde{\Gamma}(X^{0}_{\kappa}\rightarrow
K^{0}\eta)+\tilde{\Gamma}(X^{0}_{\kappa}\rightarrow K^{0}\eta')
\end{eqnarray}
where $\tilde{\Gamma}$ denotes the decay width neglecting phase
space correction({\it i.e.,} ignoring the effect of the factor
$|\vec{p}|^3$ in Eq.(\ref{17})). It can be easily checked that the
first four equations are consistent with the isospin symmetry, and
the last two equations in Eq.(\ref{20}) express the flavor cross
symmetry\cite{lipkin}. The effective lagrangian (\ref{16}) describes
the decays allowed by the OZI rule, and the contributions of the
other couplings which violate the OZI rule are neglectable in the
first order. Since $X^{\pm}_a$ and $X^{0}_a$ form a isospin triplet,
the pole position of these states should be approximately equal.
Under this approximation and using Eq.(\ref{17}), we can further
obtain the following ratio:
\begin{eqnarray}
\label{21}\Gamma(X^{+}_a\rightarrow
K^{+}K_{L}):\Gamma(X^{+}_a\rightarrow K^{+}K_{S})\approx1:1
\end{eqnarray}

We can search the other members of the nonet $X$ in $J/\psi$ decay,
{\it e.g.,} we can search $X^{+}_a$ which is the $I_3=1$ isospin
partner of $X^{0}_{a}(X(1576))$ in $J/\psi\rightarrow
X^{+}_a\pi^{-}\rightarrow K^{+}K_{L}\pi^{-}$ or $J/\psi\rightarrow
X^{+}_a\pi^{-}\rightarrow K^{+}K_{S}\pi^{-}$.
However, since $X^{0}_a(X(1576))\rightarrow K^{+}K^{-}$ has been
observed, this prediction is naturally the outcome of isospin
conservation, and any rational proposal about the nature of X(1576)
should produce this result. So this prediction can not distinguish
the different models about X(1576), and we should search some
particular signals which are almost unique in our model. With this
idea in mind, we will investigate another strong decay mode
$X\rightarrow pseudoscalar+vector$. Generally the width of these
resonances is very large, so it is likely that some members of the
vector nonet disappear into the continuum and can not be observed.
\subsection{$X\rightarrow$ pseudoscalar+vector}
The OZI allowed decays can be described by the effective lagrangian:
\begin{eqnarray}
\nonumber{\cal
L}_{eff}&=&g_2\;\varepsilon^{\mu\nu\alpha\beta}\{(X^{+}_{a})_{\mu\nu}[\rho^{-}_{\alpha\beta}\eta_s+\phi_{\alpha\beta}\pi^{-}-K^{*-}_{\alpha\beta}K^{0}-K^{*0}_{\alpha\beta}K^{-}]\\
\nonumber&&+(X^{-}_{a})_{\mu\nu}[\rho^{+}_{\alpha\beta}\eta_s+\phi_{\alpha\beta}\pi^{+}-K^{*+}_{\alpha\beta}\overline{K}^{0}-\overline{K}^{*0}_{\alpha\beta}K^{+}]\\
\nonumber&&+(X^{+}_{\kappa})_{\mu\nu}[-\rho^{-}_{\alpha\beta}\overline{K}^{0}-\overline{K}^{*0}_{\alpha\beta}\pi^{-}+\frac{1}{\sqrt{2}}K^{*-}_{\alpha\beta}(-\pi^{0}+\eta_q)+\frac{1}{\sqrt{2}}(-\rho^{0}_{\alpha\beta}+\omega_{\alpha\beta})K^{-}]\\
\nonumber&&+(X^{-}_{\kappa})_{\mu\nu}[-\rho^{+}_{\alpha\beta}K^{0}-K^{*0}_{\alpha\beta}\pi^{+}+\frac{1}{\sqrt{2}}K^{*+}_{\alpha\beta}(-\pi^{0}+\eta_q)+\frac{1}{\sqrt{2}}(-\rho^{0}_{\alpha\beta}+\omega_{\alpha\beta})K^{+}]\\
\nonumber&&+(X^{0}_{\kappa})_{\mu\nu}[-K^{*-}_{\alpha\beta}\pi^{+}-\rho^{+}_{\alpha\beta}K^{-}+\frac{1}{\sqrt{2}}(\rho^{0}_{\alpha\beta}+\omega_{\alpha\beta})\overline{K}^{0}+\frac{1}{\sqrt{2}}\overline{K}^{*0}_{\alpha\beta}(\pi^{0}+\eta_{q})]\\
\nonumber&&+(\overline{X}^{0}_{\kappa})_{\mu\nu}[-K^{*+}_{\alpha\beta}\pi^{-}-\rho^{-}_{\alpha\beta}K^{+}+\frac{1}{\sqrt{2}}(\rho^{0}_{\alpha\beta}+\omega_{\alpha\beta})K^{0}+\frac{1}{\sqrt{2}}K^{*0}_{\alpha\beta}(\pi^{0}+\eta_{q})]\\
\nonumber&&+(X^{0}_{a})_{\mu\nu}\frac{1}{\sqrt{2}}[-K^{*-}_{\alpha\beta}K^{+}-K^{*+}_{\alpha\beta}K^{-}+\overline{K}^{*0}_{\alpha\beta}K^{0}+K^{*0}_{\alpha\beta}\overline{K}^{0}+\sqrt{2}\rho^{0}_{\alpha\beta}\eta_s+\sqrt{2}\phi_{\alpha\beta}\pi^{0}]\\
\nonumber&&+(X^{0}_{f})_{\mu\nu}\frac{1}{\sqrt{2}}[K^{*-}_{\alpha\beta}K^{+}+K^{*+}_{\alpha\beta}K^{-}+\overline{K}^{*0}_{\alpha\beta}K^{0}+K^{*0}_{\alpha\beta}\overline{K}^{0}-\sqrt{2}\omega_{\alpha\beta}\eta_s-\sqrt{2}\phi_{\alpha\beta}\eta_q]\\
\label{21}&&+(X^{0}_{\sigma})_{\mu\nu}[\rho^{-}_{\alpha\beta}\pi^{+}+\rho^{+}_{\alpha\beta}\pi^{-}+\rho^{0}_{\alpha\beta}\pi^{0}-\omega_{\alpha\beta}\eta_q]\}
\end{eqnarray}
where $(X^{+}_{a})_{\mu\nu}$ is the field strength, and it is
defined by
$(X^{+}_{a})_{\mu\nu}=\partial_{\mu}(X^{+}_{a})_{\nu}-\partial_{\nu}(X^{+}_{a})_{\mu}$,
the meanings of $(X^{0}_{a})_{\mu\nu}$, $\rho^{+}_{\alpha\beta}$
$etc$ are similar. The dimension of the constant $g_2$ is
$(\rm{mass})^{-1}$($[g_2]=\rm{M}^{-1}$). Generally the decay width
is
\begin{equation}
\label{22}\Gamma(X\rightarrow
P+V)=\frac{4g^2_2}{3\pi}D_{X\rightarrow
PV}\frac{2}{(1-\beta)\sqrt{2\pi}\;\Gamma_X}\int^{m_X+\delta}_{m_X-\delta}dm\;|\vec{p}|^3\exp[-\frac{(m-m_X)^2}{2(\Gamma_X/2)^2}]
\end{equation}
Here $|\vec{p}|$ is the momentum of the vector meson $V$ or that of
the pseudoscalar $P$,
$|\vec{p}|=\frac{\sqrt{(m^2-(m_{\rm{P}}+m_{\rm{V}})^2)(m^2-(m_{\rm{P}}-m_{\rm{V}})^2)}}{2m}$,
$\delta=1.64\frac{\Gamma_X}{2}$, $\beta=10\%$\cite{pdg}.
$m_{\rm{P}}$, $m_{\rm{V}}$ are respectively the mass of the
pseudoscalar $P$ and the vector meson $V$. Being similar to
$C_{X\rightarrow P_1P_2}$, $D_{X\rightarrow PV}$ is also a numerical
coefficient, which can be read from the lagrangian (\ref{21}), and
$D_{X\rightarrow PV}$ for various decay channels are listed in Table
II.
\begin{table}[hptb]
\begin{center}
\caption{the numerical coefficient $D_{X\rightarrow PV}$ entering
Eq.(\ref{22}) for the decay of the nonet X}
\begin{tabular}{|c|c|c|c|}\hline\hline
Decay  &   $D_{X\rightarrow PV}$  &  Decay  &  $D_{X\rightarrow
PV}$\\ \hline

$X^{+}_a\rightarrow K^{*+}\overline{K}^{0}$ & 1 &
$X^{+}_{a}\rightarrow \overline{K}^{*0}K^{+}$ & 1
\\ \hline

$X^{+}_a\rightarrow \rho^{+}\eta$ &
$(\sqrt{\frac{1}{3}}\sin\theta_p-\sqrt{\frac{2}{3}}\cos\theta_p)^2$
& $X^{+}_{a}\rightarrow \rho^{+}\eta'$ &
$(\sqrt{\frac{1}{3}}\cos\theta_p+\sqrt{\frac{2}{3}}\sin\theta_p)^2$
\\ \hline

$X^{+}_a\rightarrow \phi\pi^{+}$ & 1 & $X^{+}_{\kappa}\rightarrow
\rho^{+}K^{0}$ & 1 \\ \hline

$X^{+}_{\kappa}\rightarrow K^{*0}\pi^{+}$ & 1
&$X^{+}_{\kappa}\rightarrow K^{*+}\pi^{0}$ & $\frac{1}{2}$\\ \hline

$X^{+}_{\kappa}\rightarrow K^{*+}\eta$ &
$\frac{1}{2}(\sqrt{\frac{2}{3}}\sin\theta_p+\sqrt{\frac{1}{3}}\cos\theta_p)^2$&
$X^{+}_{\kappa}\rightarrow K^{*+}\eta'$ &
$\frac{1}{2}(\sqrt{\frac{2}{3}}\cos\theta_p-\sqrt{\frac{1}{3}}\sin\theta_p)^2$\\
\hline

$X^{+}_{\kappa}\rightarrow \rho^{0}K^{+}$ & $\frac{1}{2}$
&$X^{+}_{\kappa}\rightarrow \omega K^{+}$ & $\frac{1}{2}$\\ \hline

$X^{0}_{\kappa}\rightarrow K^{*+}\pi^{-}$ & 1
&$X^{0}_{\kappa}\rightarrow \rho^{-}K^{+}$ & 1\\ \hline

$X^{0}_{\kappa}\rightarrow \rho^{0}K^{0}$ & $\frac{1}{2}$
&$X^{0}_{\kappa}\rightarrow \omega K^{0}$ & $\frac{1}{2}$\\ \hline

$X^{0}_{\kappa}\rightarrow K^{*0}\pi^{0}$ & $\frac{1}{2}$
&$X^{0}_{\kappa}\rightarrow K^{*0}\eta$ &
$\frac{1}{2}(\sqrt{\frac{2}{3}}\sin\theta_p+\sqrt{\frac{1}{3}}\cos\theta_p)^2$\\
\hline

$X^{0}_{\kappa}\rightarrow K^{*0}\eta'$ &
$\frac{1}{2}(\sqrt{\frac{2}{3}}\cos\theta_p-\sqrt{\frac{1}{3}}\sin\theta_p)^2$
& $X^{0}_{a}\rightarrow K^{*+}K^{-}$&$\frac{1}{2}$\\ \hline

$X^{0}_{a}\rightarrow K^{*-}K^{+}$&$\frac{1}{2}$
&$X^{0}_{a}\rightarrow K^{*0}\overline{K}^{0}$&$\frac{1}{2}$\\
\hline

$X^{0}_{a}\rightarrow \overline{K}^{*0}K^{0}$&$\frac{1}{2}$ &
$X^{0}_{a}\rightarrow \rho^{0}\eta$
&$(\sqrt{\frac{1}{3}}\sin\theta_p-\sqrt{\frac{2}{3}}\cos\theta_p)^2$\\
\hline

$X^{0}_{a}\rightarrow \rho^{0}\eta'$
&$(\sqrt{\frac{1}{3}}\cos\theta_p+\sqrt{\frac{2}{3}}\sin\theta_p)^2$&
$X^{0}_{a}\rightarrow \phi\pi^{0}$&1\\ \hline

$X^{0}_{f}\rightarrow
K^{*+}K^{-}$&$\frac{1}{2}$&$X^{0}_{f}\rightarrow
K^{*-}K^{+}$&$\frac{1}{2}$\\ \hline

$X^{0}_{f}\rightarrow K^{*0}\overline{K}^{0}$&$\frac{1}{2}$&
$X^{0}_{f}\rightarrow \overline{K}^{*0}K^{0}$&$\frac{1}{2}$\\ \hline

$X^{0}_{f}\rightarrow \omega\eta$
&$(\sqrt{\frac{1}{3}}\sin\theta_p-\sqrt{\frac{2}{3}}\cos\theta_p)^2$&
$X^{0}_{f}\rightarrow \omega\eta'$
&$(\sqrt{\frac{1}{3}}\cos\theta_p+\sqrt{\frac{2}{3}}\sin\theta_p)^2$\\
\hline

$X^{0}_{f}\rightarrow \phi\eta$
&$(\sqrt{\frac{2}{3}}\sin\theta_p+\sqrt{\frac{1}{3}}\cos\theta_p)^2$&
$X^{0}_{f}\rightarrow \phi\eta'$
&$(\sqrt{\frac{2}{3}}\cos\theta_p-\sqrt{\frac{1}{3}}\sin\theta_p)^2$\\
\hline

$X^{0}_{\sigma}\rightarrow \rho^{+}\pi^{-}$ &1&
$X^{0}_{\sigma}\rightarrow \rho^{-}\pi^{+}$ &1\\ \hline

$X^{0}_{\sigma}\rightarrow \rho^{0}\pi^{0}$& 1&
$X^{0}_{\sigma}\rightarrow
\omega\eta$ & $(\sqrt{\frac{2}{3}}\sin\theta_p+\sqrt{\frac{1}{3}}\cos\theta_p)^2$\\
\hline

$X^{0}_{\sigma}\rightarrow
\omega\eta'$&$(\sqrt{\frac{2}{3}}\cos\theta_p-\sqrt{\frac{1}{3}}\sin\theta_p)^2$&
& \\ \hline\hline
\end{tabular}
\end{center}
\end{table}

The decay channels which can be obtained from the channels appearing
in Table II by making charge conjugation, are not shown. From this
table, we can learn that $X^{0}_{a}(X(1576))$ can decay into
$K^{*+}K^{-}$, $K^{*-}K^{+}$, $K^{*0}K_{L}$, $K^{*0}K_{S}$,
$\overline{K}^{*0}K_{L}$, $\overline{K}^{*0}K_{S}$, $\rho^{0}\eta$,
$\rho^{0}\eta'$, $\phi\pi^{0}$. Since the pole position of
$X^{0}_{a}(X(1576))$ is bellow the threshold of $\rho^{0}\eta'$, the
process $X^{0}_{a}(X(1576))\rightarrow\rho^{0}\eta'$ only occurs
from the tail of its mass distribution. Some interesting relations
can be obtained,
\begin{eqnarray}
\nonumber&&\Gamma(X^{+}_a\rightarrow
K^{*+}\overline{K}^{0})\approx\Gamma(X^{+}_a\rightarrow
\overline{K}^{*0}K^{+})\approx2\Gamma(X^{0}_a\rightarrow K^{*+}K^{-})\approx2\Gamma(X^{0}_f\rightarrow K^{*+}K^{-})\\
\nonumber&&\Gamma(X^{0}_{a}\rightarrow
K^{*+}K^{-})\approx\Gamma(X^{0}_{a}\rightarrow
K^{*-}K^{+})\approx\Gamma(X^{0}_{a}\rightarrow
K^{*0}\overline{K}^{0})\approx\Gamma(X^{0}_{a}\rightarrow
\overline{K}^{*0}K^{0})\\
\nonumber&&\Gamma(X^{0}_{f}\rightarrow
K^{*+}K^{-})\approx\Gamma(X^{0}_{f}\rightarrow
K^{*-}K^{+})\approx\Gamma(X^{0}_{f}\rightarrow
K^{*0}\overline{K}^{0})\approx\Gamma(X^{0}_{f}\rightarrow
\overline{K}^{*0}K^{0})\\
\nonumber&&\Gamma(X^{+}_{a}\rightarrow
\rho^{+}\eta)\approx\Gamma(X^{0}_{a}\rightarrow
\rho^{0}\eta)\approx\Gamma(X^{0}_{f}\rightarrow \omega\eta)\\
\nonumber&&\Gamma(X^{+}_{a}\rightarrow
\rho^{+}\eta')\approx\Gamma(X^{0}_{a}\rightarrow
\rho^{0}\eta')\approx\Gamma(X^{0}_{f}\rightarrow \omega\eta')\\
\label{23}&&\Gamma(X^{+}_{a}\rightarrow
\phi\pi^+)\approx\Gamma(X^{0}_{a}\rightarrow \phi\pi^0)
\end{eqnarray}
\begin{eqnarray}
\nonumber&&\Gamma(X^{+}_{\kappa}\rightarrow\rho^{+}K^{0})\approx\Gamma(X^{0}_{\kappa}\rightarrow\rho^{-}K^{+})\\
\nonumber&&\Gamma(X^{+}_{\kappa}\rightarrow
\rho^{0}K^{+})\approx\Gamma(X^{+}_{\kappa}\rightarrow
\omega K^{+})\approx\frac{1}{2}\Gamma(X^{+}_{\kappa}\rightarrow\rho^{+}K^{0})\\
\nonumber&&\Gamma(X^{0}_{\kappa}\rightarrow\rho^{0}K^{0})\approx\Gamma(X^{0}_{\kappa}\rightarrow\omega
K^{0})\approx\frac{1}{2}\Gamma(X^{0}_{\kappa}\rightarrow\rho^{-}K^{+})\\
\nonumber&&\Gamma(X^{+}_{\kappa}\rightarrow
K^{*0}\pi^{+})\approx2\Gamma(X^{+}_{\kappa}\rightarrow
K^{*+}\pi^{0})\approx\Gamma(X^{0}_{\kappa}\rightarrow
K^{*+}\pi^{-})\approx2\Gamma(X^{0}_{\kappa}\rightarrow
K^{*0}\pi^{0})\\
\nonumber&&\Gamma(X^{+}_{\kappa}\rightarrow
K^{*+}\eta)\approx\Gamma(X^{0}_{\kappa}\rightarrow
K^{*0}\eta),~~~\Gamma(X^{+}_{\kappa}\rightarrow
K^{*+}\eta')\approx\Gamma(X^{0}_{\kappa}\rightarrow K^{*0}\eta')\\
\label{24}&&\Gamma(X^{0}_{\sigma}\rightarrow\rho^{-}\pi^{+})\approx\Gamma(X^{0}_{\sigma}\rightarrow
\rho^{+}\pi^{-})\approx\Gamma(X^{0}_{\sigma}\rightarrow
\rho^{0}\pi^{0})
\end{eqnarray}
\begin{eqnarray}
\nonumber&&\tilde{\Gamma}(X^{+}_{a}\rightarrow
K^{*+}\overline{K}^{0})+\tilde{\Gamma}(X^{+}_{a}\rightarrow
\overline{K}^{*0}K^{+})\approx\tilde{\Gamma}(X^{+}_{a}\rightarrow
\rho^{+}\eta)+\tilde{\Gamma}(X^{+}_{a}\rightarrow
\rho^{+}\eta')+\tilde{\Gamma}(X^{+}_{a}\rightarrow \phi\pi^{+})\\
\nonumber&&\tilde{\Gamma}(X^{0}_{a}\rightarrow
K^{*+}K^{-})+\tilde{\Gamma}(X^{0}_{a}\rightarrow
K^{*-}K^{+})+\tilde{\Gamma}(X^{0}_{a}\rightarrow
K^{*0}\overline{K}^{0})+\tilde{\Gamma}(X^{0}_{a}\rightarrow\overline{K}^{*0}K^{*0})\\
\nonumber&&\approx\tilde{\Gamma}(X^{0}_{a}\rightarrow\rho^{0}\eta)+\tilde{\Gamma}(X^{0}_{a}\rightarrow\rho^{0}\eta')+\tilde{\Gamma}(X^{0}_{a}\rightarrow\phi\pi^{0})\\
\nonumber&&\tilde{\Gamma}(X^{+}_{\kappa}\rightarrow
\rho^{+}K^{0})+\tilde{\Gamma}(X^{+}_{\kappa}\rightarrow
K^{*0}\pi^{+})\approx\tilde{\Gamma}(X^{+}_{\kappa}\rightarrow
K^{*+}\pi^{0})+\tilde{\Gamma}(X^{+}_{\kappa}\rightarrow K^{*+}\eta)\\
\nonumber&&+\tilde{\Gamma}(X^{+}_{\kappa}\rightarrow
K^{*+}\eta')+\tilde{\Gamma}(X^{+}_{\kappa}\rightarrow\rho^{0}K^{+})+\tilde{\Gamma}(X^{+}_{\kappa}\rightarrow\omega
K^{+})\\
\nonumber&&\tilde{\Gamma}(X^{0}_{f}\rightarrow
K^{*+}K^{-})+\tilde{\Gamma}(X^{0}_{f}\rightarrow
K^{*-}K^{+})+\tilde{\Gamma}(X^{0}_{f}\rightarrow
K^{*0}\overline{K}^{0})+\tilde{\Gamma}(X^{0}_{f}\rightarrow
\overline{K}^{*0}K^{0})\\
\label{26}&&\approx\tilde{\Gamma}(X^{0}_{f}\rightarrow
\omega\eta)+\tilde{\Gamma}(X^{0}_{f}\rightarrow
\omega\eta')+\tilde{\Gamma}(X^{0}_{f}\rightarrow
\phi\eta)+\tilde{\Gamma}(X^{0}_{f}\rightarrow \phi\eta'),
\end{eqnarray}
where $\tilde{\Gamma}$ denotes the partial decay width neglecting
phase space. We can see that Eq.(\ref{23}) and Eq.(\ref{24}) are
consistent with the isospin symmetry. The first equation in
Eq(\ref{26}) is exactly the Eq.(4) of the Ref.\cite{lipkin}, and the
equations in Eq.(\ref{26}) reflect the flavor cross symmetry in the
decay of the four quark states. Using Eq.(\ref{22}) and the pole
position of $X(1576)$:
$(1576^{+49+98}_{-55-91})$MeV-i$(409^{+11+32}_{-12-67})\rm{MeV}$, we
can further obtain,
\begin{eqnarray}
\nonumber&&\Gamma(X^+_{a}(X(1576))\rightarrow
K^{*+}\overline{K}^{0}):\Gamma(X^+_{a}(X(1576))\rightarrow
\overline{K}^{*0}K^{+}):\Gamma(X^+_{a}(X(1576))\rightarrow
\rho^{+}\eta)\\
\nonumber&&:\Gamma(X^+_{a}(X(1576))\rightarrow
\rho^{+}\eta'):\Gamma(X^+_{a}(X(1576))\rightarrow
\phi\pi^{+})\approx1:1:0.47:0.175:1.24\\
\nonumber&&\Gamma(X^0_{a}(X(1576))\rightarrow
K^{*+}K^{-}):\Gamma(X^0_{a}(X(1576))\rightarrow
K^{*0}\overline{K}^{0}):\Gamma(X^0_{a}(X(1576))\rightarrow\rho^{0}\eta)\\
\label{27}&&:\Gamma(X^0_{a}(X(1576))\rightarrow\rho^{0}\eta'):\Gamma(X^0_{a}(X(1576))\rightarrow\phi\pi^{0})\approx1:1:0.94:0.35:2.48
\end{eqnarray}
The above ratios shows that in our four quark state scenario, the
decay $X^0_{a}(X(1576))\rightarrow\phi\pi^{0}$ is favorable, which
is the distinctive feature of our four quark state interpretation.
We expect the $I_3=1$ state $X^{+}_{a}$ should also appear in
$J/\psi\rightarrow X^{+}_{a}\pi^{-}\rightarrow\phi\pi^{+}\pi^{-}$ ,
$J/\psi\rightarrow X^{+}_{a}\pi^{-}\rightarrow
\overline{K}^{*0}K^{+}\pi^{-}$ and $J/\psi\rightarrow
X^{+}_{a}\pi^{-}\rightarrow K^{*+}\overline{K}^{0}\pi^{-}$ .
Experimental search of the channel $X^0_{a}(X(1576))\rightarrow
pseudoscalar+vector$ is necessary so that the existence of X(1576)
can be reexamined.
\section{conclusion and discussion}
We propose that $X(1576)$ recently reported by BES collaboration can
be interpreted as the diquark-antidiquark bound state in P-wave
excitation. This implies that there exists a vector nonet $X$, and
$X(1576)$ is a member of the nonet. We estimate the mass spectrum of
the nonet by considering both the colormagnetic hyperfine
interaction energy and the P-wave excitation energy. The theoretical
prediction for the mass of $X(1576)$ is 1632.854MeV, which is
consistent with the experimental data:
$(1576^{+49+98}_{-55-91})$MeV-i$(409^{+11+32}_{-12-67})$MeV. The
strong coupling of X(1576) to its decay channel $K^{+}K^{-}$ may
affect both the imaginary part of the pole position and its real
part\cite{fsi}, this effect is ignored in the work, which is need to
be studied further. Because the experimental error on the pole
position is large, we expect the prediction will also be consistent
with the experimental data if this effect is taken into account. The
diquark here is taken as "good" diquark, generally the "bad" diquark
is involved. However, the lowest lying and more stable states are
dominated by the "good" diquark configuration\cite{jaffe1,jaffe2}.
Dealing with the mixing effects exactly from quark model is in
progress.

OZI allowed strong decay of the nonet are investigated in detail.
Both two pseudoscalars decay channel and one pseudoscalar plus one
vector meson channel are discussed. We find out that in our four
quark state scheme, the dominant decay modes of $X^{0}_a(X(1576))$
are $K^{+}K^{-}$, $K_{L}K_{S}$, $\phi\pi^{0}$, but not
$\pi^{+}\pi^{-}$, and this is a important test for our proposal. We
predict that the positive and negative charged isospin partner of
$X^{0}_a(X(1576))$ dominantly decay into strange mesons. Since these
two states are connected by charge conjugation, we concentrate on
the positive charged $I_3=1$ states $X^{+}_a$. In order to search
these states, we suggest to analyze the $J/\psi$ decay data in
$J/\psi\rightarrow
K^{+}K_{L}\pi^{-}$, $J/\psi\rightarrow K^{+}K_{S}\pi^{-}$ and
$J/\psi\rightarrow \phi\pi^{+}\pi^{-}$. The observation of $X^{+}_a$
is another crucial test of our scheme. The decays of the other
members of the nonet are also discussed, which can provide important
clue to the experimental search of these states. Similar to
$X^{0}_a$ (i.e., $X(1576))$, the width of these states should be
broad too, and hence it is also difficult to observe them
experimentally.

Diquark is in $\mathbf{3}_c$ configuration, so diquark and
antidiquark can not be observed individually. As the distance
between diquark and antidiquark gets large, a $\rm{q}\overline{q}$
will be created from the vacuum, then the state decays into
baryon-antibaryon. But the central values of the mass distribution
of the nonet are generally bellow the threshold of the
baryon-antibaryon pair, the decay width should be small.

These states can mix with the ordinary $\rm{q\bar{q}}$ states, if
they have the same quantum numbers ( {\it e.g.,} $X^{0}_a(X(1576))$
can mix with $\rho(1450)$, $\rho(1700)$ and so on ). The mixing
effects interfere in the spectrum and the decay properies, and a
full consideration including mixing effects is notoriously a
difficult problem in exotic hadron spectrum. More sophisticated
treatment of $X(1576)$ which takes these effects into account is
expected. However, as a first step to understand $X(1576)$, we
expect that such mixing effects in most cases are small from
previous work on multiquark states\cite{jaffe1,jaffe2}, and the
results obtained in this letter are at least correct qualitatively.

Recently BES performs a partial wave analysis of
$J/\psi\rightarrow\phi\pi^{+}\pi^{-}$ and $J/\psi\rightarrow\phi
K^{+}K^{-}$ from a sample of 58M $J/\psi$ events in the BESII
detector. There is a strong peak in the $\phi\pi$ mass distribution
which centers at 1500MeV$/\rm{c}^2$ with a full-width of
200MeV$/\rm{c}^2$\cite{bes1}, and this peak was also reported about
twenty years ago with quantum number $J^{P}=1^{-}$\cite{russ}. It is
very likely there is some component of $X^{+}_a$ in this $\phi\pi$
peak. Finally, we would like to mention that if our predictions are
not consistent with future experimental results, $X(1576)$ should
have a different structure. More experimental facts about X(1576)
are needed in order to clarify these issues.

\section *{ACKNOWLEDGEMENTS}
\indent This work is partially supported by National Natural Science
Foundation of China under Grant Numbers 90403021, and by the PhD
Program Funds 20020358040 of the Education Ministry of China and
KJCX2-SW-N10 of the Chinese Science Academy.



\begin{thebibliography}{99}
\bibitem{bes}BES Collaboration, M.~Ablikim {\it et al.},
Submitted to  Phys.\ Rev.\ Lett.\ [arXiv:hep-ex/0606047].
\bibitem{gs}F.~K.~Guo and P.~N.~Shen,[arXiv:hep-ph/0606273].
\bibitem{lipkin} M.~Karliner and H.~J.~Lipkin, [arXiv:hep-ph/0607093].
\bibitem{maiani1} L.~Maiani, F.~Piccinini, A.~D.~Polosa and V.~Riquer,
Phys.\ Rev.\ D {\bf 71}, 014028 (2005)[arXiv:hep-ph/0412098].
\bibitem{maiani2}L.~Maiani, V.~Riquer, F.~Piccinini and A.~D.~Polosa,
Phys.\ Rev.\ D {\bf 72}, 031502 (2005)[arXiv:hep-ph/0507062].
\bibitem{maiani3}L.~Maiani, F.~Piccinini, A.~D.~Polosa and V.~Riquer,
Phys.\ Rev.\ Lett.\  {\bf 93}, 212002 (2004)[arXiv:hep-ph/0407017].
\bibitem{jaffe1}R.~L.~Jaffe, Phys.\ Rept.\  {\bf 409}, 1 (2005)[Nucl.\ Phys.\ Proc.\ Suppl.\  {\bf 142}, 343 (2005)]
[arXiv:hep-ph/0409065]; F.~E.~Close,Int.\ J.\ Mod.\ Phys.\ A {\bf
20}, 5156 (2005)[arXiv:hep-ph/0411396].
\bibitem{colormag}A.~De Rujula, H.~Georgi and S.~L.~Glashow,Phys.\ Rev.\ D {\bf 12}, 147 (1975).
\bibitem{lipkin2}M.~Karliner and H.~J.~Lipkin,Phys.\ Lett.\ B {\bf 575}, 249 (2003)
[arXiv:hep-ph/0402260].
\bibitem{sorba}H.~Hogaasen, J.~M.~Richard and P.~Sorba,Phys.\ Rev.\ D {\bf 73}, 054013 (2006)
[arXiv:hep-ph/0511039].
\bibitem{hs1} H.~Hogaasen and P.~Sorba,Mod.\ Phys.\ Lett.\ A {\bf 19}, 2403 (2004)[arXiv:hep-ph/0406078].
\bibitem{hs2}H.~Hogaasen and P.~Sorba,Nucl.\ Phys.\ B {\bf 145}, 119 (1978).
\bibitem{jaffe2}R.~L.~Jaffe and K.~Johnson, Phys.\ Lett.\ B {\bf 60}, 201 (1976);R.~L.~Jaffe, Phys.\ Rev.\ D {\bf 15}, 267 (1977); R.~L.~Jaffe,Phys.\ Rev.\ D {\bf 15}, 281 (1977).
\bibitem{mix}A.~Bramon, R.~Escribano and M.~D.~Scadron,Phys.\ Lett.\ B {\bf 403}, 339 (1997)[arXiv:hep-ph/9703313].
\bibitem{pdg}Particle Data Group, S.~Eidelman {\it et al.,} Phys.\ Lett.\ B{\bf 592}, 1, 2004.
\bibitem{fsi}T.~Barnes, F.~E.~Close and H.~J.~Lipkin,
Phys.\ Rev.\ D {\bf 68}, 054006 (2003)[arXiv:hep-ph/0305025]; E.~van
Beveren and G.~Rupp, Phys.\ Rev.\ Lett.\  {\bf 91}, 012003
(2003)[arXiv:hep-ph/0305035]; D.~S.~Hwang and D.~W.~Kim, Phys.\
Lett.\ B {\bf 601}, 137 (2004)[arXiv:hep-ph/0408154].
\bibitem{bes1}BES Collaboration, M.~Ablikim {\it et al.}, Phys.\ Lett.\ B {\bf 607}, 243 (2005)[arXiv:hep-ex/0411001].
\bibitem{russ}S.~I.~Bityukov {\it et al.}, Sov.\ J.\ Nucl.\ Phys.\  {\bf 38}, 727 (1983)
[Yad.\ Fiz.\  {\bf 38}, 1205 (1983)].
\end{thebibliography}
\end{document}